\documentstyle[prl,aps,floats,epsf]{revtex}

\makeatletter
\newenvironment{ifnosubmit}{\endgroup}{\begingroup\def\@currenvir{ifnosubmit}}

\makeatother
\begin{ifnosubmit}
\newenvironment{widetable}{\csname table*\endcsname}{\csname endtable*\endcsname}
\end{ifnosubmit}

\def\tt{t\bar t}
\def\pp{p\bar p}

\def\roots{\sqrt{s}}
\def\gammax{\Gamma_X}

\def\met{\mbox{${\hbox{$E$\kern-0.6em\lower-.1ex\hbox{/}}}_T$}}
\def\metcal{\mbox{${\hbox{$E$\kern-0.6em\lower-.1ex\hbox{/}}}_T$}^{cal}}

\def\metvec{\vec{\met}}

\def\pznu{p^{\nu}_z}

\newcommand\degree{^\circ}                      

\def\gevcc{{\rm GeV}/c^2}
\def\chisq{\chi^2}

\newcommand\progname[1]{{\sc\lowercase{#1}}}
\newcommand\eqref[1]{Eq.~(\ref{#1})}

\newcommand\figref[1]{Fig.~\ref{#1}}
\newcommand\tabref[1]{Table~\ref{#1}}

\newcommand\mx{{\bf x}}
\newcommand\mG{{\bf G}}
\def\chisq{\chi^2}

\long\def\simplex#1#2#3#4{
\begin{figure}[hbt]
\vspace*{-0.5cm}
\centerline{\epsfxsize=#1\epsfbox{#2}}
\caption{#3}
\label{#4}
\end{figure}
}

\begin{document}
\title{Search for Narrow $\tt$ Resonances in $\pp$ Collisions at $\roots$ = 1.8 TeV.}

%
\author{                                                                      
V.M.~Abazov,$^{21}$                                                           
B.~Abbott,$^{55}$                                                             
A.~Abdesselam,$^{11}$                                                         
M.~Abolins,$^{48}$                                                            
V.~Abramov,$^{24}$                                                            
B.S.~Acharya,$^{17}$                                                          
D.L.~Adams,$^{53}$                                                            
M.~Adams,$^{35}$                                                              
S.N.~Ahmed,$^{20}$                                                            
G.D.~Alexeev,$^{21}$
% ...
\\                                                                            
\vskip 0.30cm                                                                 
\centerline{(D\O\ Collaboration)}                                             
\vskip 0.30cm                                                                 
}                                                                             
\address{                                                                     
\centerline{$^{1}$Universidad de Buenos Aires, Buenos Aires, Argentina}       
\centerline{$^{2}$LAFEX, Centro Brasileiro de Pesquisas F{\'\i}sicas,         
                  Rio de Janeiro, Brazil}                                     
\centerline{$^{3}$Universidade do Estado do Rio de Janeiro,                   
                  Rio de Janeiro, Brazil}                                     
\centerline{$^{4}$Institute of High Energy Physics, Beijing,                  
                  People's Republic of China}                                 
\centerline{$^{5}$Universidad de los Andes, Bogot\'{a}, Colombia}             
\centerline{$^{6}$Charles University, Center for Particle Physics,            
                  Prague, Czech Republic}                                     
\centerline{$^{7}$Institute of Physics, Academy of Sciences, Center           
                  for Particle Physics, Prague, Czech Republic}               
\centerline{$^{8}$Universidad San Francisco de Quito, Quito, Ecuador}         
\centerline{$^{9}$Laboratoire de Physique Subatomique et de Cosmologie,       
                  IN2P3-CNRS, Universite de Grenoble 1, Grenoble, France}     
\centerline{$^{10}$CPPM, IN2P3-CNRS, Universit\'e de la M\'editerran\'ee,     
                  Marseille, France}                                          
\centerline{$^{11}$Laboratoire de l'Acc\'el\'erateur Lin\'eaire,              
                  IN2P3-CNRS, Orsay, France}                                  
\centerline{$^{12}$LPNHE, Universit\'es Paris VI and VII, IN2P3-CNRS,         
                  Paris, France}                                              
\centerline{$^{13}$DAPNIA/Service de Physique des Particules, CEA, Saclay,    
                  France}                                                     
\centerline{$^{14}$Universit{\"a}t Mainz, Institut f{\"u}r Physik,            
                  Mainz, Germany}                                             
\centerline{$^{15}$Panjab University, Chandigarh, India}                      
\centerline{$^{16}$Delhi University, Delhi, India}                            
\centerline{$^{17}$Tata Institute of Fundamental Research, Mumbai, India}     
\centerline{$^{18}$CINVESTAV, Mexico City, Mexico}                            
\centerline{$^{19}$FOM-Institute NIKHEF and University of                     
                  Amsterdam/NIKHEF, Amsterdam, The Netherlands}               
\centerline{$^{20}$University of Nijmegen/NIKHEF, Nijmegen, The               
                  Netherlands}                                                
\centerline{$^{21}$Joint Institute for Nuclear Research, Dubna, Russia}       
\centerline{$^{22}$Institute for Theoretical and Experimental Physics,        
                   Moscow, Russia}                                            
\centerline{$^{23}$Moscow State University, Moscow, Russia}                   
\centerline{$^{24}$Institute for High Energy Physics, Protvino, Russia}       
\centerline{$^{25}$Lancaster University, Lancaster, United Kingdom}           
\centerline{$^{26}$Imperial College, London, United Kingdom}                  
\centerline{$^{27}$University of Arizona, Tucson, Arizona 85721}              
\centerline{$^{28}$Lawrence Berkeley National Laboratory and University of    
                  California, Berkeley, California 94720}                     
\centerline{$^{29}$University of California, Davis, California 95616}         
\centerline{$^{30}$California State University, Fresno, California 93740}     
\centerline{$^{31}$University of California, Irvine, California 92697}        
\centerline{$^{32}$University of California, Riverside, California 92521}     
\centerline{$^{33}$Florida State University, Tallahassee, Florida 32306}      
\centerline{$^{34}$Fermi National Accelerator Laboratory, Batavia,            
                   Illinois 60510}                                            
\centerline{$^{35}$University of Illinois at Chicago, Chicago,                
                   Illinois 60607}                                            
\centerline{$^{36}$Northern Illinois University, DeKalb, Illinois 60115}      
\centerline{$^{37}$Northwestern University, Evanston, Illinois 60208}         
\centerline{$^{38}$Indiana University, Bloomington, Indiana 47405}            
\centerline{$^{39}$University of Notre Dame, Notre Dame, Indiana 46556}       
\centerline{$^{40}$Iowa State University, Ames, Iowa 50011}                   
\centerline{$^{41}$University of Kansas, Lawrence, Kansas 66045}              
\centerline{$^{42}$Kansas State University, Manhattan, Kansas 66506}          
\centerline{$^{43}$Louisiana Tech University, Ruston, Louisiana 71272}        
\centerline{$^{44}$University of Maryland, College Park, Maryland 20742}      
\centerline{$^{45}$Boston University, Boston, Massachusetts 02215}            
\centerline{$^{46}$Northeastern University, Boston, Massachusetts 02115}      
\centerline{$^{47}$University of Michigan, Ann Arbor, Michigan 48109}         
\centerline{$^{48}$Michigan State University, East Lansing, Michigan 48824}   
\centerline{$^{49}$University of Nebraska, Lincoln, Nebraska 68588}           
\centerline{$^{50}$Columbia University, New York, New York 10027}             
\centerline{$^{51}$University of Rochester, Rochester, New York 14627}        
\centerline{$^{52}$State University of New York, Stony Brook,                 
                   New York 11794}                                            
\centerline{$^{53}$Brookhaven National Laboratory, Upton, New York 11973}     
\centerline{$^{54}$Langston University, Langston, Oklahoma 73050}             
\centerline{$^{55}$University of Oklahoma, Norman, Oklahoma 73019}            
\centerline{$^{56}$Brown University, Providence, Rhode Island 02912}          
\centerline{$^{57}$University of Texas, Arlington, Texas 76019}               
\centerline{$^{58}$Rice University, Houston, Texas 77005}                     
\centerline{$^{59}$University of Virginia, Charlottesville, Virginia 22901}   
\centerline{$^{60}$University of Washington, Seattle, Washington 98195}       
}                                                                             

\maketitle

\begin{abstract}
A search for narrow resonances that decay into $\tt$ pairs has been performed using $130~\rm{pb^{-1}}$ of data in the lepton$+$jets channel collected in $\pp$ collisions at $\roots$ = 1.8 TeV. There is no significant deviation observed from the standard model, and upper limits at the 95$\%$ confidence level on the product of the production cross section and branching fraction to $\tt$ are presented for narrow resonances as a function of the resonance mass $M_X$. These limits are used to exclude the existence of a leptophobic topcolor particle with mass $M_X<560~\gevcc$ and width $\gammax=0.012M_X$. \\
\end{abstract}


\twocolumn

Narrow resonances decaying to $\tt$ pairs are predicted by several theories beyond the standard model~\cite{topcolor,hillparkeharris}. For instance, one of the scenarios of the topcolor-assisted technicolor model in Ref. \cite{hillparkeharris} predicts a heavy $Z'$ boson that couples preferentially to the third quark generation, and not to leptons (leptophobic). The cross section for the $Z'$ boson in this model is large enough for it to be observed over a wide range of masses and widths in data available from the 1.8 TeV $\pp$ Tevatron Collider at the Fermi National Accelerator Laboratory. \\

 In searches for such heavy particles or resonances, we seek an excess of events beyond that predicted by the standard model in the distribution of the invariant mass of $\tt$ decay products. Previous searches at the Tevatron have limited a leptophobic $Z'$ boson to a mass higher than $480~\gevcc$~\cite{cdf}. In this paper, we describe a direct search for narrow $\tt$ resonances in the inclusive decay modes $\tt\rightarrow\ell\nu~+\geq 4~{\rm{jets}}$, where $\ell={\rm an}~{\rm electron}~(e)~{\rm~or}~{\rm a}~{\rm muon}~(\mu)$, using $130~\rm{pb^{-1}}$ of data recorded by the D$\O$ experiment from 1992 to 1996. Having observed no significant deviation from the standard model, we present model-independent 95$\%$ confidence-level (C.L.) upper limits on the product of the cross section ($\sigma_X$) and branching fraction ($B$) to $\tt$, for a narrow resonance. We also present a lower limit on the resonance mass ($M_X$) of the $Z'$ boson in a particular model~\cite{hillparkeharris}. \\

The D$\O$ detector is a multi-purpose particle detector designed to study $\pp$ collisions at the Fermilab Tevatron Collider. The detector consists of three major systems: a non-magnetic central tracking system, a uranium/liquid-argon calorimeter, and a muon spectrometer. A detailed description of the D$\O$ detector can be found in Ref.~\cite{d0detector}. \\
\begin{widetable}
\caption{Summary of event selections. Here $\met^{cal}$ is the missing transverse energy measured just in the calorimeter, $\eta^W$ is the pseudorapidity of the $W$ boson that decays leptonically, and $\Delta\phi (\met,\mu)$ is the difference in the azimuthal angle between $\met$ and the highest-$p_T$ muon.} 
\vspace*{1cm}
{\small
\begin{tabular}{ccccc} 
& $e$+$jets$ & $\mu$+$jets$ &  $e$+$jets$/$\mu$ &  $\mu$+$jets$/$\mu$\\ \hline
&&&& \\
Lepton ($l$)& $E_T^l>$20 GeV  & $p_T^l>$20 GeV/$c$  &  $E_T^l>$20 GeV &$p_T^l>$20 GeV/$c$  \\
& $|\eta|<$2 & $|\eta|<$1.7 & $|\eta|<$2 &$|\eta|<$1.7 \\\hline
$\met$&$\met >$20 GeV & $\met >$20 GeV & $\met >$20 GeV & \hspace*{-.5cm} 
$\met >$20 GeV \\
& $\met^{cal}>$25 GeV & $\met^{cal}>$20 GeV  &&$\met^{cal}>$20 GeV\\\hline
Jets& $\ge$ 4 jets & $\ge$ 4 jets & $\ge$ 4 jets &  $\ge$ 4 jets \\
& $E_T>$15 GeV & $E_T>$15 GeV & $E_T>$15 GeV &$E_T>$15 GeV \\
& $|\eta|<$2 & $|\eta|<$2 & $|\eta|<$2 & $|\eta|<$2 \\\hline
$\mu$ tag& No & No & Yes & Yes \\\hline

Other& $|\met|+|E_T^l|>$ 60 GeV & $|\met|+|p_T^l|>$ 60 GeV &$\met >$35 GeV,&$\Delta\phi (\met,\mu) 
< 170 \degree$ \\
& $|\eta^W|<$2  & $|\eta^W|<$2&if $\Delta\phi (\met,\mu) < 25\degree$&
$|\Delta\phi(\met,\mu)-90\degree|/90\degree$ \\

&  & & &$<\met/$(45 GeV) \\\hline

Events passing  &  &  &  &  \\
above criteria & 42 & 41 & 4 & 3 \\\hline
& & & & \\
With $\chisq < 10$ & 16 & 21 & 1 & 3 \\ 
\end{tabular}
}
\label{tb:prcuts}
\end{widetable}

The present search rests upon techniques developed for the measurement of the mass of the top quark at D$\O$ in the $\rm lepton+jets$ channel~\cite{topmassprd}. Due to the large mass of the top quark ($m_t$), the $\tt\rightarrow\ell\nu~+\geq 4~{\rm{jets}}$ final state is characterized by a high-$p_T$ isolated lepton ($e~{\rm or}~\mu$) and large missing transverse energy ($\met$) from the undetected neutrino. Additional soft muons ($\mu~\rm tags$) from semileptonic decays of $b$ and $c$ quarks occur in $\approx20\%$ of $\tt$ events but only in $\approx2\%$ of non-$\tt$ events~\cite{topquarksearch}, and therefore offer discrimination between signal and background. We consider two orthogonal classes of events for this analysis: a) a purely topological selection of lepton+jets events denoted as $e+jets$ and $\mu+jets$, where the jets do not contain a muon, and b) a selection based primarily on the presence of a muon contained within a jet ($\mu$ tag), and additional selections on the topology of the event. These events are denoted as $e+jets/\mu$ and $\mu+jets/\mu$. Details of the trigger requirements, reconstruction of events, and identification of the $e$, $\mu$, $\met$, and jets can be found in Ref.~\cite{topmassprd}. The principal sources of background correspond to standard-model $\tt$ production, $W (\rightarrow l\nu)+\rm jets$ production, and production of multijets $(N_{j} \approx 5)$, in which one of the jets is misidentified as a lepton and $\met$ stems from jet-energy mismeasurement. For the measurement of the top-quark mass, most selections were optimized to reduce the contribution from non-$\tt$ sources. We therefore use similar selections in the present analysis, and these are summarized in \tabref{tb:prcuts}.   \\
 
The resonance signal $X~\rightarrow~\tt$ is modeled using the \progname{pythia}-6.1~\cite{pythia} Monte Carlo event generator, with $m_t$ = 175 $\gevcc$, and CTEQ3M~\cite{pdf_cteq} parton distribution functions. Initial and final-state radiation (ISR/FSR) is included. About 10,000 events at nine resonance masses between 400 and 1000 $\gevcc$ are generated, using a width $\gammax=0.012 M_X$. This width is significantly smaller than the $\approx 0.04 M_X$ mass resolution of the D$\O$ detector for $\tt$ systems~\cite{mythesis}. Hence, our results are dominated by the detector resolution and independent of $\gammax$. The generated events are processed through the \progname{d\o geant} detector simulation package~\cite{d0geant1} and reconstructed using the D$\O$ event-reconstruction program. A standard set of corrections is applied to electromagnetic objects and jets~\cite{topmassprd}, and the missing transverse energy recalculated. \\ 

The backgrounds are estimated from a combination of Monte Carlo simulations and collider data~\cite{topmassprd}. The selections summarized in \tabref{tb:prcuts} are also applied to the Monte Carlo (MC) signal and background samples. \\

Each event in data, as well as in the Monte Carlo signal and background samples, is fitted to a three-constraint (3C) hypothesis for the $\tt$ production and decay:
\begin{eqnarray}
\label{eq:star1}
\tt \rightarrow W^+b~W^-\bar{b}, \\\nonumber
W^+ \rightarrow l^+ \nu_l ~~({\rm or} ~ q {\bar q'}), \\\nonumber
W^- \rightarrow q {\bar q'}~~ ({\rm or}~ l^- {\bar \nu_l}). \nonumber
\end{eqnarray}
The inputs to the fit are the measured kinematic parameters of the lepton and the jets, and the missing transverse energy vector, $\metvec$. We minimize $\chisq~=~(\mx - \mx^m)^T \mG (\mx - \mx^m)$, where $\mx^m(\mx)$ is the vector for measured (fitted) variables, and $ \mG^{-1}$ is its error matrix~\cite{topmassprd}. The two reconstructed $W$ boson masses are constrained to the pole mass $M_W$ of the $W$ boson, and the reconstructed $t$ and $\bar{t}$ quark masses are set to $m_t~=~173.3~\gevcc$~\cite{topmassprd}. Only the four highest-$E_T$ jets are used in the kinematic fit. All other jets are assumed to be due to initial-state radiation, and are ignored. There are 6 (12) possible assignments of these jets to quarks in the events with (without) a $\mu$ tag, each having two solutions for the longitudinal momentum of the neutrino ($\pznu$). For every possible permutation, we apply additional parton-level and $\eta$-dependent jet corrections derived using data and Monte Carlo simulations~\cite{topmassprd}. We apply a loose selection on the reconstructed mass, M($q\bar q$), of the hadronically decaying $W$ boson, $40<M(q\bar q)<140~ \gevcc$, before the fit, to reduce computation. The results of the fit with the lowest $\chisq$ are used to reconstruct the invariant mass ($M_{\tt}$) of the $\tt$ system. It is observed that the jet permutation with the lowest $\chisq$ is the correct choice for $\approx 20\%$ of all Monte Carlo $\tt$ events~\cite{topmassprd}. We require $\chisq<10$ to further reduce non-$\tt$ background, whereupon 41 events are left in the data sample, of which four are $\mu$-tagged. \\

For each $M_X$ sample generated by Monte Carlo, we perform a fit based on Bayesian statistics~\cite{bayes} to determine the number of events expected from signal and background in the observed lepton$+$jets data sample. We fit~\cite{mythesis} the data to a three-source model comprised of signal ($X~\rightarrow~\tt$), and backgrounds from standard-model $\tt$ production, $W+\rm jets$, and multijets. We combine backgrounds from $W+\rm jets$ and multijets in the ratio 0.78:0.22, based on a measurement of their relative proportions in the top-quark mass analysis at D$\O$~\cite{topmassprd}. We define a likelihood ($L$) and a posterior probability $P(n_1,n_2,n_3,M_X|D)$ for obtaining $n_1$, $n_2$ and $n_3$ events from the three respective sources, for a model specified by $M_X$. Given the observed data set $D$, we can write:

{\footnotesize   
\begin{equation}
\hspace*{-0.5mm}P(n_1,\hspace*{-0.3mm}n_2,\hspace*{-0.3mm}n_3,M_{\hspace*{-0.5mm}X}|D)\hspace*{-0.8mm}=\hspace*{-0.8mm}\frac{L(D|n_1,\hspace*{-0.3mm}n_2,\hspace*{-0.3mm}n_3,\hspace*{-0.3mm}M_{\hspace*{-0.5mm}X})w(n_1,\hspace*{-0.3mm}n_2,\hspace*{-0.3mm}n_3|M_{\hspace*{-0.5mm}X})}{\mathcal{N'}},
\label{eq:post}
\end{equation}
}
\hspace*{-1.3mm}where $w$ denotes the joint prior probability for the three source strengths, and $\mathcal{N'}$ is a normalization that is obtained from the requirement:   
\begin{equation}
\int P(n_1,n_2,n_3,M_X|D)dn_1dn_2dn_3 = 1. 
\label{eq:postzzz}
\end{equation}
We assume Poisson statistics for the likelihood, and flat priors for each of the three sources. Bayesian integration~\cite{bayes} over possible signal and background populations in each bin $i$ of the $M_{\tt}$ distribution yields the likelihood:
%
\begin{eqnarray}
\nonumber
L(D|n_1,n_2,n_3,M_X) &=& \prod_{i=1}^M \sum_{k_1,k_2,k_3=0}^{D_i}\prod_{j=1}^3
  {A_{ji} + k_j \choose k_j} \\
&& \times {p_j^{k_j} \over (1 + p_j)^{A_{ji} + k_j + 1}},  
\label{eq:likedef}
\end{eqnarray}
where $D_i$ ($A_{ji}$) is the number of events in bin $i$ for data (Monte Carlo source $j$); the indices $k_j$ satisfy the multinomial constraint $\sum_{j=1}^3 k_j = D_i$; $p_j~=~n_j/(M~+~\sum_{i=1}^MA_{ji})$ is an estimate of the strength of the $j^{th}$ source ($j=1,2,3$); and $M$ is the number of bins. The expected number of counts from any source $j$ can be obtained from the fit as:  
%
\begin{equation}
<n_j>=\int\hspace*{-1.0mm}\int\hspace*{-1.0mm}\int n_j P(n_1,n_2,n_3,M_X|D)dn_1dn_2dn_3. 
\label{eq:nexpj}
\end{equation}
The \hspace{0.55mm}fitted \hspace{0.55mm}number \hspace{0.55mm}of \hspace{0.55mm}events \hspace{0.55mm}expected \hspace{0.55mm}from \hspace{0.55mm}the \hspace{0.55mm}signal ($<n_1>$) \hspace{0.55mm}and \hspace{0.55mm}the \hspace{0.55mm}two \hspace{0.55mm}background \hspace{0.55mm}sources $(<n_{2}> {\rm and} <n_{3}>)$ are listed in \tabref{tb:normcounts} for several values of $M_X$. The observed $M_{\tt}$ distribution and the corresponding distributions from the three \hspace{0.55mm}Monte \hspace{0.55mm}Carlo \hspace{0.55mm}sources \hspace{0.55mm}normalized \hspace{0.55mm}to $<n_1>$, \hspace{0.55mm}$<n_2>$ \hspace{0.55mm}and \hspace{0.55mm}$<n_3>$, \hspace{0.55mm}respectively, \hspace{0.55mm}for $M_X=400~\gevcc$, are shown in \figref{norm-mtt}. There is no significant deviation from the standard-model prediction. Similar agreement is observed for other choices of resonance mass. \\
\begin{table}[h]
\begin{center}
\caption{\hspace*{0.8mm}The fitted number of events expected from signal, $<n_1>$, and background from standard model $\tt$ production, $<n_{2}>$, and $W+\rm jets$ and multijets, $<n_{3}>$, for different $M_X$. After all selections, 41 events are observed in the $M_{\tt}$ distribution of lepton+jets data. } 
\vspace*{1.cm}
\begin{tabular}{c|cccc} 
$M_X$ & $<n_1>$ & $<n_2>$ & $<n_3>$ & Background \\ 
(GeV/$c^2$)&&&&{\hspace*{-0.3cm}\small $<n_2>$+$<n_3>$} \\\hline
400 &  9.0$\pm$7.0 & 20.5$\pm$10.8 & 13.9$\pm$10.2 & 34.4$\pm$14.9 \\ 
500 &  4.9$\pm$4.2 & 22.2$\pm$11.5 & 15.3$\pm$10.5 & 37.5$\pm$15.6 \\  
600 &  4.2$\pm$3.2 & 23.7$\pm$11.6 & 15.4$\pm$10.6 & 39.0$\pm$15.7 \\ 
750 &  1.6$\pm$1.6 & 26.8$\pm$11.7 & 12.6$\pm$9.9 & 39.4$\pm$15.3 \\ 
\end{tabular}
\label{tb:normcounts}
\end{center}
\end{table}
\simplex{8.0cm}{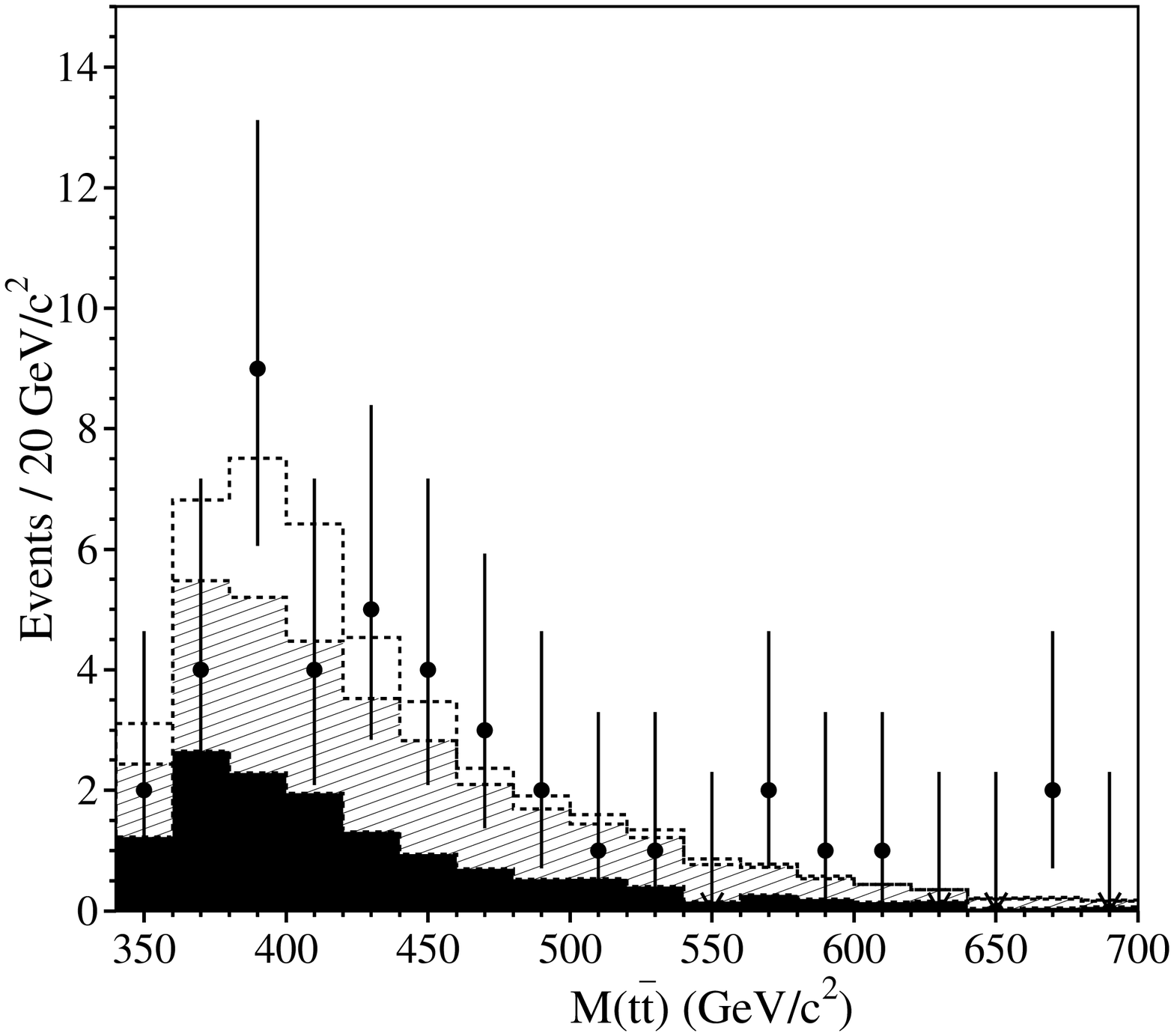}{\footnotesize Distributions of $M_{\tt}$ obtained from the fit, for the sum of signal ($X\rightarrow\tt$) and all standard-model backgrounds (open histogram), sum of all standard-model backgrounds (hatched histogram), and $W$+jets and multijets (solid histogram), for $M_X=400~\gevcc$. The data correspond to the dots with their statistical errors.}{norm-mtt}

In the absence of a signal, we proceed to set upper limits on the product of the production cross section of $X$ and branching fraction to $\tt$, $\sigma_XB$, by expressing $n_1={\mathcal{AL}}\sigma_X B$ in \eqref{eq:post}, where $\mathcal{A}$ is the acceptance for $X\rightarrow\tt$ events and $\mathcal{L}$ is the integrated luminosity. Integrating over $n_2$ and $n_3$, we define for every $M_X$ the upper limit on $\sigma_X B$ at the $95\%$ confidence level as:
%
\begin{equation}
\int_0^{(\sigma_X B)_{95}}P(\sigma_X B, M_X|D) d(\sigma_X B) = 0.95.
\label{eq:limit95}
\end{equation}
%

The expected shapes of distributions for background and signal, and the acceptance for signal, are subject to several sources of systematic uncertainty. The uncertainty due to the jet energy scale is estimated by re-scaling the jet energies by $\pm(2.5\%+0.5~ \rm GeV)$~\cite{topmassprd} before applying any selections to the signal Monte-Carlo events. For the contribution from ISR/FSR, we compare the acceptance for the signal with and without ISR/FSR (in \progname{pythia}). For the uncertainty from the choice of parton distribution functions, we compare the signal acceptance for the two parton distribution sets CTEQ3M and GRV94L~\cite{grv94l}. We also consider the uncertainties in trigger efficiency, lepton identification, and integrated luminosity. All the sources of statistical and systematic uncertainty in the product $\mathcal{AL}$ are listed in \tabref{tb:mxerrors} for $M_X=400~\gevcc$~\cite{mythesis}. 
\begin{table}[h]
\vspace*{0.2cm}
\begin{center}
\caption{The fractional uncertainty in the product $\mathcal{AL}$ from different sources, for $M_X~=~400~\gevcc$.} 
\vspace*{0.5cm}
\begin{tabular}{cc} 
      MC statistics          &  3.3 $\%$\\
      Trigger efficiency     &  3.6 $\%$\\
      $e/\mu$ identification &  3.8 $\%$\\
      Luminosity             &  4.3 $\%$\\
      Jet energy scale       &  7.4 $\%$\\
      ISR/FSR                & 16.0 $\%$\\
      PDF                    & 15.0 $\%$\\
      \hline
     Total                  & 24.3 $\%$\\
\end{tabular}
\label{tb:mxerrors}
\end{center}
\end{table}

For each $M_X$, we convolute the posterior probability density $P(\sigma_X B, M_X|D)$ with a Gaussian prior for $\mathcal{A}\mathcal{L}$, with the estimated value of $\mathcal{A}\mathcal{L}$ as the mean of the Gaussian and its uncertainty as one standard deviation from the mean. The upper limits on $\sigma_X B$ at the 95$\%$ confidence level obtained using \eqref{eq:limit95}, integrating over all possible values of $\mathcal{A}\mathcal{L}$, are listed in \tabref{tb:results}. We use these limits to constrain~\cite{mythesis} a model of topcolor-assisted technicolor, and exclude at the 95$\%$ C.L. the existence of a leptophobic $Z'$ boson with mass $M_X < 560~\gevcc$, for a width $\gammax=0.012M_X$, as shown in \figref{mxbound}. \\ 

\begin{table}[h]
\vspace*{0.2cm}
\begin{center}
\caption{The 95$\%$ C.L. upper limits on $\sigma_XB$ for narrow resonances of mass $M_X$ decaying into $\tt$.} 
\vspace*{0.5cm}
\begin{tabular}{cc} 
$M_{X}$& 95$\%$ C.L. upper limits on \\
(GeV/$c^2$)&$\sigma_X B$ (pb) \\ \hline
400.&5.0 \\
450&4.5 \\
500&2.7  \\
550&2.3 \\
600&2.3 \\
650&2.0 \\
750&1.3 \\
850&1.5 \\
1000&2.0 \\
\end{tabular}
\label{tb:results}
\end{center}
\end{table}
\simplex{8.0cm}{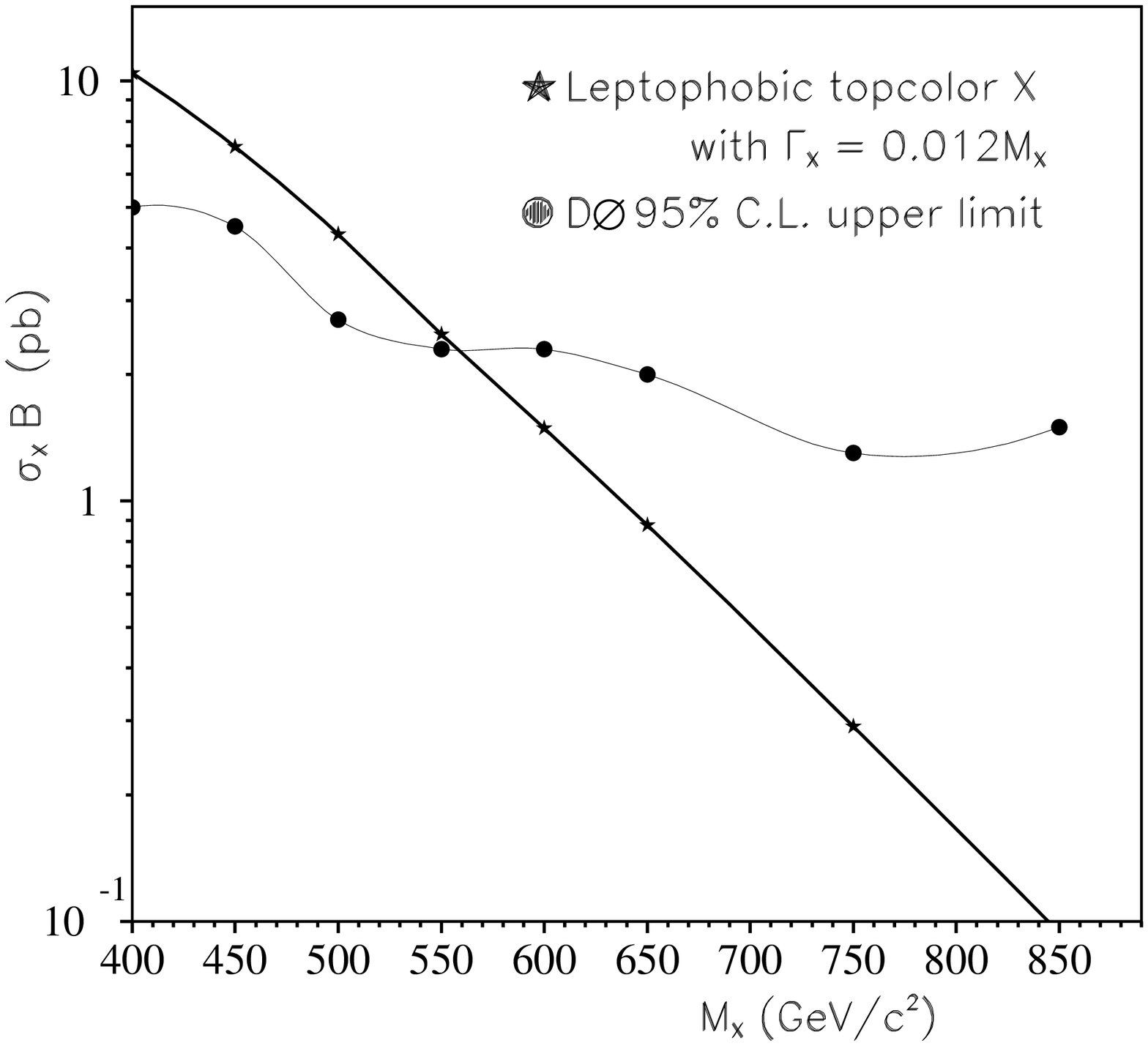}{\footnotesize The 95$\%$ C.L. upper limit on $\sigma_X B$ as a function of resonance mass $M_X$. Included for reference is the predicted topcolor-assisted technicolor cross section for a width $\gammax=0.012M_X$.}{mxbound}
 
In conclusion, after investigating 130 $\rm{pb^{-1}}$ of data, we find no statistically significant evidence for a $\tt$ resonance, and establish upper limits on $\sigma_X B$ at the 95$\%$ C.L. for $M_X$ between 400 and 1000 $\gevcc$. We also exclude at the 95$\%$ C.L. the existence of a leptophobic $Z'$ boson with mass $M_X < 560~\gevcc$, for a width $\gammax=0.012M_X$. \\
%
%

We thank the staffs at Fermilab and collaborating institutions, 
and acknowledge support from the 
Department of Energy and National Science Foundation (USA),  
Commissariat  \` a L'Energie Atomique and 
CNRS/Institut National de Physique Nucl\'eaire et 
de Physique des Particules (France), 
Ministry for Science and Technology and Ministry for Atomic 
   Energy (Russia),
CAPES, CNPq and FAPERJ (Brazil),
Departments of Atomic Energy and Science and Education (India),
Colciencias (Colombia),
CONACyT (Mexico),
Ministry of Education and KOSEF (Korea),
CONICET and UBACyT (Argentina),
The Foundation for Fundamental Research on Matter (The Netherlands),
PPARC (United Kingdom),
Ministry of Education (Czech Republic),
A.P.~Sloan Foundation,
and the Research Corporation.
%

\end{document}